%% file: main.tex
\documentclass[%
 reprint,
 amsmath,amssymb,
prstab,
]{revtex4-2}

\usepackage{graphicx}
\usepackage{dcolumn}
\usepackage{bm}
\usepackage[ruled,vlined,linesnumbered]{algorithm2e}
\usepackage{mathtools}

\usepackage{xcolor}

\begin{document}


\title[RCDS-S]{An optimization method to compensate accelerator performance drifts}

\author{Zhe Zhang}
\author{Minghao Song}
 \altaffiliation[Also at ]{Illinois Institute of Technology, Chicago, IL 60616.}
\author{Xiaobiao Huang}%
 \email{xiahuang@slac.stanford.edu}
\affiliation{%
 SLAC National Accelerator Laboratory, Menlo Park, CA 94025
}%

\date{\today}

\begin{abstract}
Accelerator performance often deteriorates with time during a long period of operation due to 
secular changes in the machine components or the surrounding environment. 
In many cases some tuning knobs are effective in compensating the performance drifts and optimization methods can be used to find the ideal machine setting. However, 
such intervention usually cannot be done without interrupting user operation 
as the optimization algorithms can substantially impact the machine performance.
We propose an optimization algorithm, Safe Robust Conjugate Direction Search (RCDS-S), 
which can perform accelerator tuning while keeping the machine performance within a 
designated safe envelope. The algorithm builds probability models of the objective function using Lipschitz continuity of the function as well as characteristics of the drifts and applies to  the selection of trial solutions to ensure the machine operates safely during tuning. 
The algorithm can run during normal user operation constantly, or periodically, to compensate the performance drifts. 
Simulation and online tests have been done to validate the performance of the algorithm. 
\end{abstract}

\maketitle

\section{Introduction}

Online optimization is an effective approach to find accelerator settings with high performance. 
In online optimization, a number of machine control parameters (i.e., tuning knobs) are varied by an optimization algorithm to minimize or maximize the objective function, which represents the machine performance and is measured experimentally for 
each machine setting. Efficient optimization algorithms are key to online optimization. 
Popular optimization algorithms for online accelerator applications include Nelder-Mead simplex~\cite{NelderMead1965}, robust conjugate direction search (RCDS)~\cite{HUANG201377}, particle swarm~\cite{KennedyPSO1995}, and Bayesian optimization~\cite{Duris2020PRL}. 

Typically, when the algorithm suggests new trial solutions, the step size from known solutions is not restricted and the performance of the trial solution is not guaranteed. 
During an optimization run, as the algorithm gradually discovers machine settings with high performance, it can also produce solutions with poor performance, which cannot be tolerated for normal user operation. 
Therefore, online optimization is usually performed during dedicated machine development or study shifts. 
Often times, after the ideal machine setting is found, it is delivered to user operation for days or weeks, until the next window for tuning becomes available. 

However, in many cases, an ideal machine setting will not maintain the high performance during the long period of user operation. Small variations in the accelerator components, caused by or coupled with variations of the surrounding environment, can cause the machine performance to drift with time. The underlying causes of the performance drift are not always known; even if a connection with certain environmental factors can be established, the relationship is often not adequately deterministic to build reliable feedforward control. 
On the other hand, the performance drift can usually be compensated by tuning some control knobs. 
Unfortunately, as such tuning can only be done during dedicated shifts, one has to either tolerate the deteriorating machine performance or interrupt the operation schedule to add a tuning session. 

In this study, we propose a safe tuning method that can be used during user operation. The new algorithm is called safe robust conjugate direction search (RCDS-S). It employs iterative one-dimensional (1-D) optimization over a conjugate direction set in a similar manner as the RCDS method. However, its 1-D optimization is done by a more prudent and informed fashion, which employs a probability model of the objective function to assess the risk of exceeding a safety threshold by the trial solution. The probability model includes the combined effect of the innate variation of the objective function and the slow drift with time.

The proposed method is similar to an earlier method \cite{sui2015safe, kirschner2019adaptive}, in that both studies make use of the Lipschitz continuity properties of the objective function. However, there are significant differences between the two methods. In Ref.~\cite{kirschner2019adaptive},  a safe domain is identified through  the Lipschitz continuous condition, over which Gaussian process optimization is performed. In our method, the Lipschitz continuity condition is used to select trial solutions based on the need to explore the parameter space and the safety requirement;  the trial solutions are used in a parabolic fitting to determine the location of the minimum. 

The new method has been successfully tested with a real-life problem in both simulations and experiments. The test problem is  kicker bump matching of a storage ring. In the tests, the amplitude of a kicker is modulated independently, while the amplitudes of the other two kickers are tuned by the algorithm. It was shown that the objective function can be kept below a pre-specified threshold during the tunig period. A beam steering for optimal injection efficiency problem is also used to test the algorithm in simulation. 

This paper is organized as follows. Section~\ref{secRCDSS}  describes the RCDS-S algorithm. 
Section~\ref{secSimul} shows the application of RCDS-S to two simulated drifting accelerator problems. 
Section~\ref{secExpe} presents experimental results when the method is applied to a real 
accelerator. Section~\ref{secAlt} discusses an alternative approach to model the system drift and shows a few preliminary test results. 
Section~\ref{secConclu} gives the conclusion.

\section{The RCDS-S method\label{secRCDSS}}

Our goal of the study is to develop an optimization method that can be used to optimize accelerator performance during user operation by keeping the performance above a certain threshold. Such a method could be termed ``safe'' optimization algorithm. A safe optimization algorithm could be used to compensate the performance drift with time, as it can run in the background continuously,  periodically, or as needed. 
To achieve the goal, it is necessary to first understand the sources of the danger in the usual ``unsafe'' methods, before a cure can be found. 

In the following, we first discuss the uncertainty of the objective function as it is probed. By constructing a probability model of the uncertainty and using it to guide the selection of new trial solutions, we devised a safe 1-D optimization method. Combining this safe 1-D optimization method and the conjugate direction search method, we arrived at the  new algorithm, RCDS-S. 

\subsection{Modeling uncertainty of objective function}\label{ssec:drift_model_a}

In this study, we cast an  optimization problem as to minimize the objective function with a set of tuning knobs, $f({\bf x})$, where elements of ${\bf x}$ are the knob values. An online  optimization problem has measurement errors, therefore,
\[
y = f(\mathbf{x}) + \epsilon,
\]
where $\epsilon \sim N(0, \sigma_n^2)$, $\sigma_n$ is the standard deviation of measurement errors.

In  online optimization, the optimization algorithm continues to sample the objective function by evaluating new trial solutions. The  sampling results are used by the algorithm to determine the next trial solution. Depending on the efficiency of the algorithm, the overall trend among all the trial solutions is  to improve the objective function. However, for  the traditional algorithms,  a trial solution at any time could correspond to a poor performance, as they typically do not limit the step size with the predicted performance. 
In this case, the intrinsic variation of the objective function in the parameter space poses as a danger. 

To be able to suggest a safe trial solution, the algorithm has to have a sort of model about the unknown objective function. Our approach is to simply limit the gradient of the function. Mathematically, 
the objective function is assumed to be L-Lipschitz continuous,
which means for any $\mathbf{x}, \mathbf{x_0} \in \mathbf{D}$, 
where $\mathbf{D}$ is the domain of the function, we have, 
\[
\left\|f(\mathbf{x}) - f(\mathbf{x_0})\right\| \leq L\cdot\left\|\mathbf{x}-\mathbf{x_0}\right\|.
\]
Considering noise, it leads to
\[
y \leq y_0 + L\cdot\left\|\mathbf{x}-\mathbf{x_0}\right\| + \sqrt{\sigma^2 + \sigma_0^2}\cdot\hat{\epsilon},
\]
were $\hat{\epsilon} \sim N(0, 1)$. 
Clearly, the expectation of $y$ satisfies
\[
E(y) \leq y_0 + L\cdot\left\|\mathbf{x}-\mathbf{x_0}\right\| = E_{\max}(\mathbf{x}). 
\]

The goal of the safety search method is to keep the objective function below a certain safety threshold, for any trial solutions the algorithm proposes during the optimization. 
Let the safety threshold be $h$, point $\mathbf{x}$ would be a safe point, i.e., 
$y \leq h$,
if
\begin{equation}
\hat{\epsilon} \leq \frac{h - E_{\max}}{\sqrt{\sigma^2 + \sigma_0^2}}.
\label{equ:safety_cond}
\end{equation}
The probability for Eq.~\eqref{equ:safety_cond} to hold for point $\mathbf{x}$ is
\begin{equation}
p(\mathbf{x}) = \frac{1}{2}\cdot\left[1+\textrm{erf}\left(\frac{h - E_{\max}(\mathbf{x})}{\sqrt{2(\sigma^2 + \sigma_0^2)}}\right)\right], 
\label{equ:safety_prob}
\end{equation}
where $\textrm{erf}(\cdot)$ is the error function. 

A second source of danger in online tuning is the  
 drift, or the time-dependent variation, of the objective function. 
Without further information about the specific optimization problem, the drift can be modeled as a random walk process.
Under this assumption, the uncertainty of the measurement becomes a time varying random variable, 
\[
y = f(\mathbf{x}) + \epsilon(t).
\]

When the drift at one point $\mathbf{x}$ is treated as a Gaussian random walk process, the measured function value differs from the original value by a random amount, whose standard deviation increases with time according to 
\[
\sigma_t = \sqrt{t}\cdot\sigma_d,
\]
where $\sigma_d$ is the drift rate,
with  $\sigma_d^2$ representing the increase of the variance within a unit time interval, and $t$ the time elapsed from a reference point. It would be reasonable to assume that noise and drift are not coupled, therefore $\epsilon(t) \sim N(0, \sigma_n^2 + \sigma_t^2 )$.
Combined with Eq.~\eqref{equ:safety_cond},
we have
\begin{equation}
\hat{\epsilon} \leq \frac{h - E_{\max}}{\sqrt{2\sigma_n^2 + t\cdot\sigma_d^2}},
\label{equ:drift_a}
\end{equation}
which is used to update Eq.~\eqref{equ:safety_prob}.

\subsection{1-D safety exploration without drift}

By applying the probability model of the objective function to the 1-D subspace, i.e., along a line in the parameter space, we can develop a strategy to safely explore the linear space. 
The idea is that we calculate the safety probability for all  points (i.e., potential trial solutions) along the full viable range, using each previous observation point based on Eq.~\eqref{equ:safety_prob}.
For any candidate point, we take the maximum of all the calculated probabilities as the estimated safety probability for that point. For example,
if 5 observations are given, we have 5 probabilities, $p_1\sim p_5$, calculated from the observations at position $x$, and the final assigned safety probability of $x$ is $p(x) = \max(p_1, p_2, p_3, p_4, p_5)$. In this way, we obtain a safety probability curve along the line.

\begin{figure}[htbp]
\begin{center}
\includegraphics[width=\linewidth]{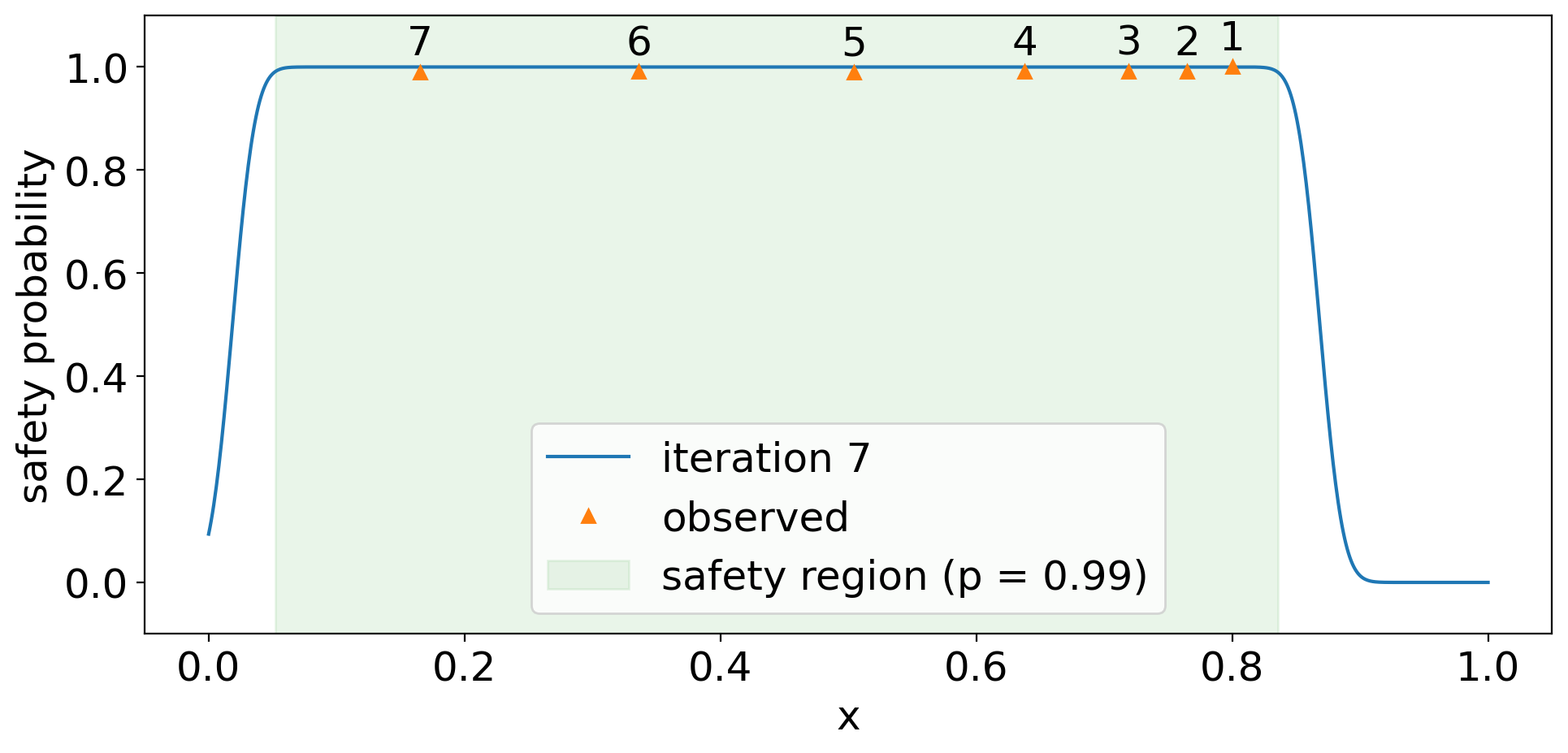}
\caption{Snapshot of the safety probability along the viable range at the end of 1-D safety exploration. Safety probability curve is calculated with the 7 given observations. The order of sampling is indicated by the number on top of each observation. Orange triangles denote the observations, blue curve shows the calculated safety probability, the safety region with safety probability threshold $p_s=0.99$ is highlighted in green.}\label{fig:safety_prob}
\end{center}
\end{figure}

To determine the next point to sample, we introduce a performance indicator (PI): safety probability threshold $p_s$, which indicates the required safety probability for safe exploration.
For example, $p_s=0.99$ means the algorithm should only choose the point as the candidate where the assigned safety probability is higher than 0.99. With safety probability threshold defined, the safety region can be computed.

To illustrate the 1-D safety exploration strategy, we introduce a simple test function, 
\begin{equation}
y = f(x) + \epsilon = C(x - \mu)^2 + \sigma_n\hat{\epsilon},
\label{eq:test_1d}
\end{equation}
where $C = L/(2\max(\mu, 1-\mu))$, $\mu \in [0, 1]$, $x \in [0, 1]$, $\sigma_n$ the standard deviation of a Gaussian noise, $\hat{\epsilon}$ a random variable that obeys the standard normal distribution. This test function $f(x)$ is  L-Lipschitz continuous. In this case study, we choose Lipschitz constant $L=1$ and safety threshold $h=0.2$.
Fig.~\ref{fig:safety_prob} shows an example of the safety probability curve and the safety region (highlighted in green), when 7 observation points are given. 

Similar to the RCDS method, the purpose of line exploration is to collect enough data points to bracket the minimum and to perform a parabola fitting. Before the minimum is bracketed, it would be preferred to expand the sample coverage region as far as possible. Therefore, the next data point is chosen to be one of two edge points of the safety region, usually the one further away from the known observation points.

\begin{figure}[htbp]
\begin{center}
\includegraphics[width=\linewidth]{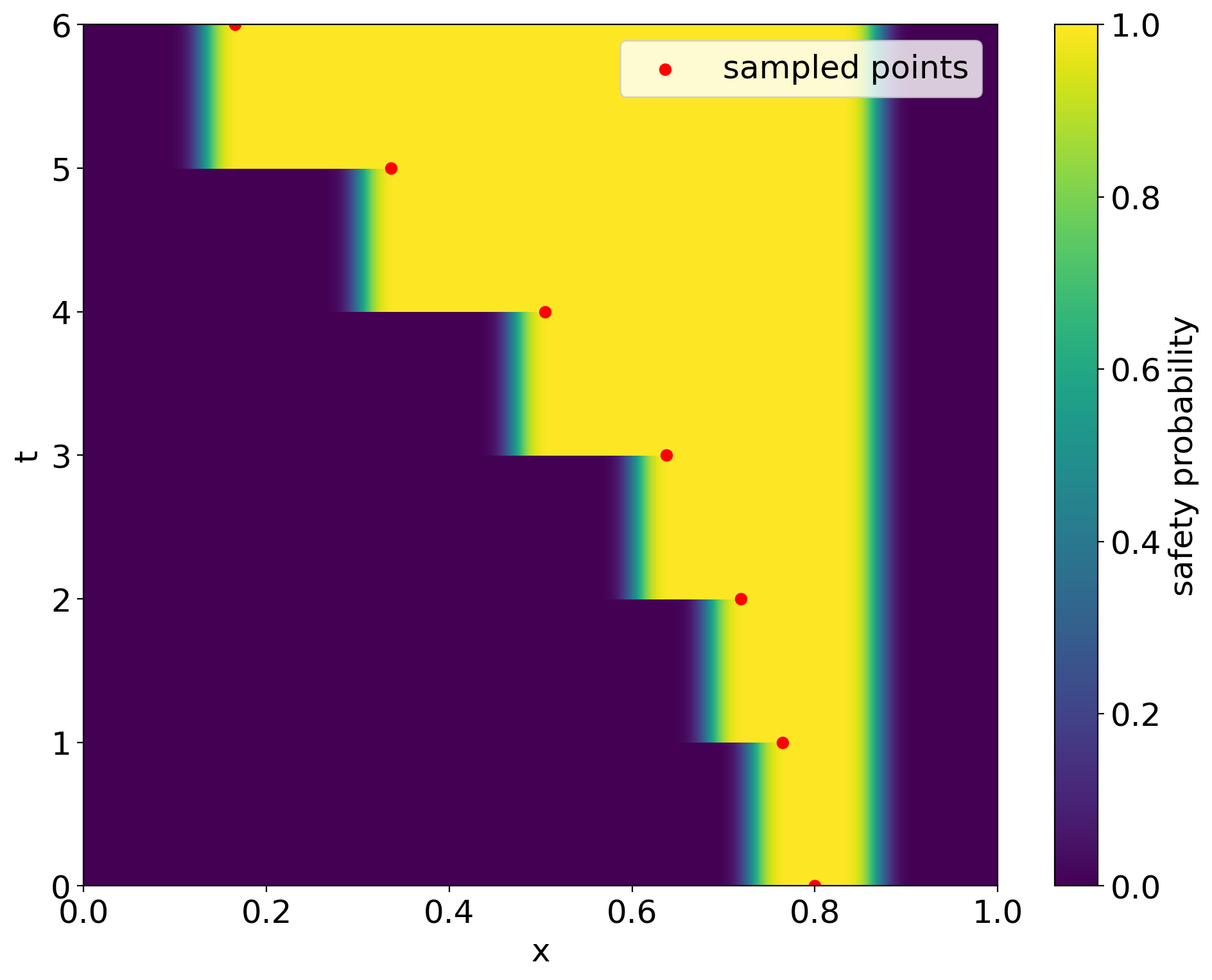}
\caption{Evolution of the safety probability during a safety exploration on the 1-D test problem. Red dot denotes the time and position of the corresponding sampled point.
}\label{fig:static_model}
\end{center}
\end{figure}

\begin{figure}[htbp]
\begin{center}
\includegraphics[width=\linewidth]{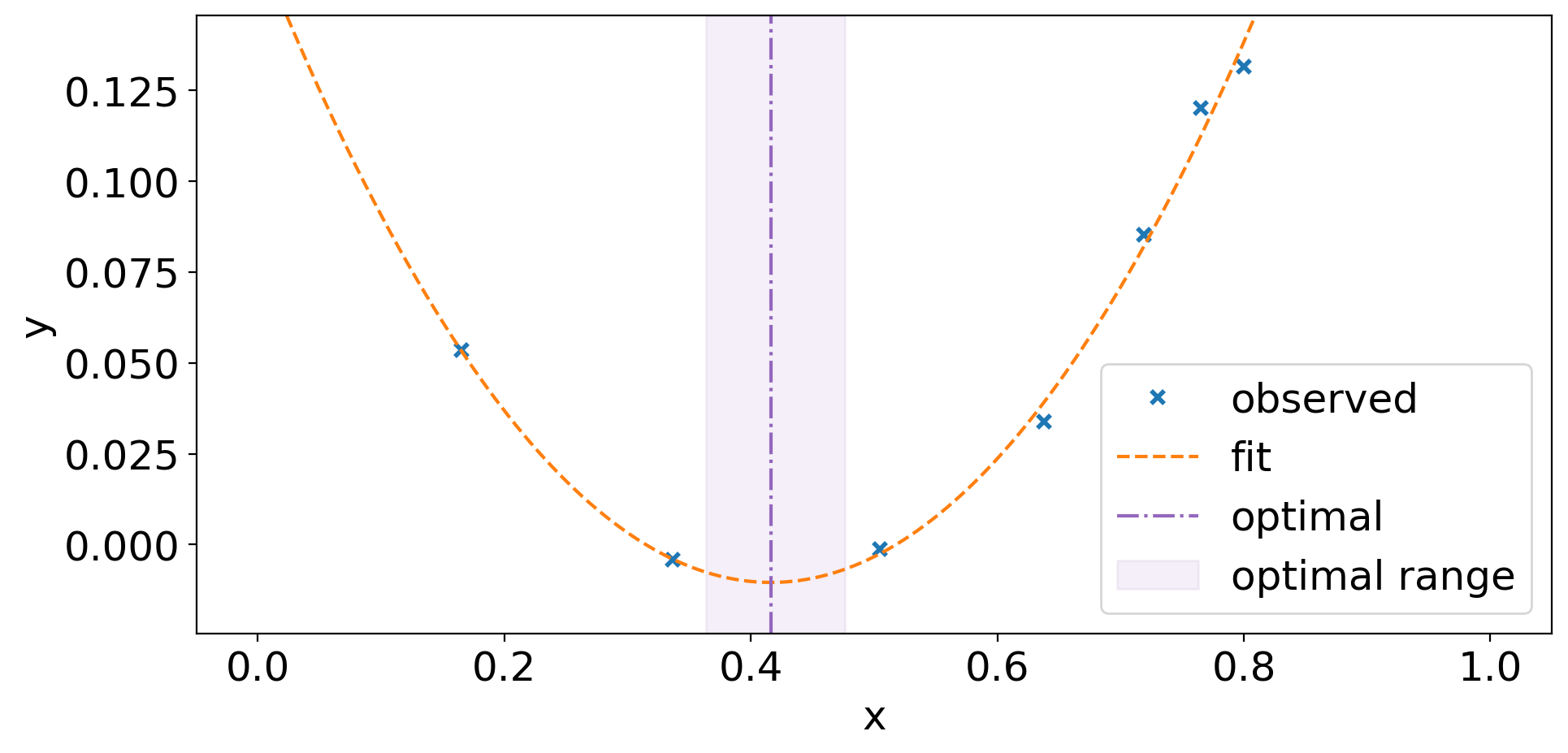}
\caption{Parabola fitting at the end of exploration. The purple dashed vertical line indicates the peak position found by the fitting, the purple region around the peak line shows the uncertainty of the peak position, calculated by the covariance matrix of the parabola fitting.}\label{fig:para_fit}
\end{center}
\end{figure}

When the trial solution is evaluated, it becomes an observation point. 
The exploration process described above repeats until we have enough data points to do a parabola fitting (as shown in Fig.~\ref{fig:para_fit}). More detailed logic and the implementation of safety exploration along one direction can be found in Algorithm~\ref{alg:linescan}. 
The safety exploration algorithm has three exit conditions, which indicate the exploration is successful, aborted as no safe candidates are available, or the maximum number of trial solutions are attempted but without success, respectively. 

\IncMargin{0.4em}
\begin{algorithm}
\SetAlgoLined
 $n\gets 0$, $\mathcal{S}_0 \gets [x_0]$, $\mathcal{O}_0 \gets [y_0]$. Initialize $\mathcal{C}$\;
 \While{$n < N_{max}$}{
  Check if the peak has already been bracketed, and
  try to perform parabola fitting with $\mathcal{S}_n$ and $\mathcal{O}_n$\;
  \eIf{peak is bracketed \textbf{and} parabola fitting succeeded}
  {
    Terminate the exploration. Report error code 0, peak position $x_p$\;
  }{
    For each candidate in $\mathcal{C}$, calculate the safety probability $p$ based on $\mathcal{S}_n$ and $\mathcal{O}_n$\;
    $\mathcal{C}_s \gets$ candidates with safety probability $p > p_{\tiny{\textrm{thres}}}$\;
    \While{$\textrm{range}(\mathcal{C}_s) \in \textrm{range}(\mathcal{S}_n)$}{
      Lower down the safety probability threshold $p_{\tiny{\textrm{thres}}}$\;
      \If{$p_{\tiny{\textrm{thres}}}$ is below a lower limit $\epsilon$} {
        Give up the exploration. Report error code 1\;
      }
      $\mathcal{C}_s \gets$ candidates with safety probability $p > p_{\tiny{\textrm{thres}}}$\;
    }
    Select candidate from $\mathcal{C}_s$ that has maximum distance from $\mathcal{S}_n$ as the next position-to-sample $x_{n+1}$\;
    Evaluate $x_{n+1}$ to get $y_{n+1}$\;
    $\mathcal{S}_{n+1} \gets \mathcal{S}_{n} + [x_{n+1}]$, $\mathcal{O}_{n+1} \gets \mathcal{O}_{n} + [y_{n+1}]$\;
    $n\gets n+1$\;
  }
 }
 Report error code -1\;
 \caption{1D Safety Exploration}
 \label{alg:linescan}
\end{algorithm}
\DecMargin{0.4em}

\subsection{1-D safety exploration with drift}\label{sec:drift_1d}

As discussed earlier, when drift to the objective function is introduced, the calculation of the safety probability changes (see Eq.~\eqref{equ:drift_a}). 
Intuitively, an older observations would become less reliable than the newer observations. As a consequence, the contribution of each observation to the safety probability calculation  differ with measurement time.
The 1-D safety exploration procedure is applicable to the case with drift, only with the changes to the safety probability calculation. 

We will use the same test problem, shown in Eq.~\eqref{eq:test_1d}, for illustrating and testing, but with modulation to the minimum position $\mu$ to simulate the systematic drift, 
\[
\mu(t) = \mu_0 + \sigma_d\sin\left(\frac{2\pi t}{p}\right),
\]
where $p$ and  $\sigma_d$ are the period and amplitude of the modulation, respectively. 

\begin{figure}[htbp]
\begin{center}
\includegraphics[width=\linewidth]{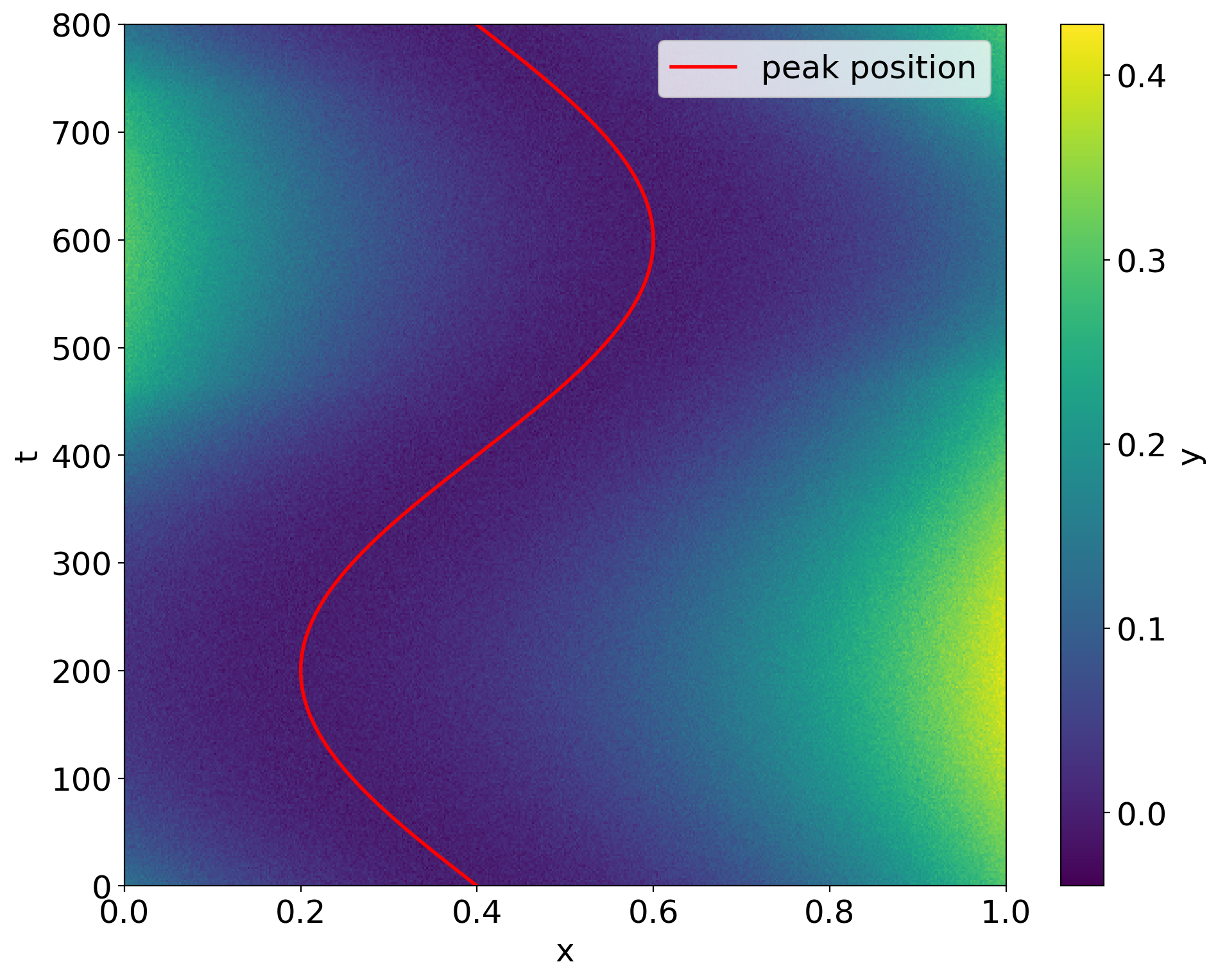}
\caption{Test function surface evolution with time. The drifting period $p$ is set to 800, and drifting amplitude $\sigma_d$ is 0.2. The noise level $\sigma_n$ is 0.01. Red curve is the peak position of the test function that changes with time.
}\label{fig:drift_surface}
\end{center}
\end{figure}

The  test function with drift, $f(x, t)$, is shown in Fig~\ref{fig:drift_surface}, where the red line indicates the drifting of the minimum position. 

\begin{figure}[htbp]
\begin{center}
\includegraphics[width=\linewidth]{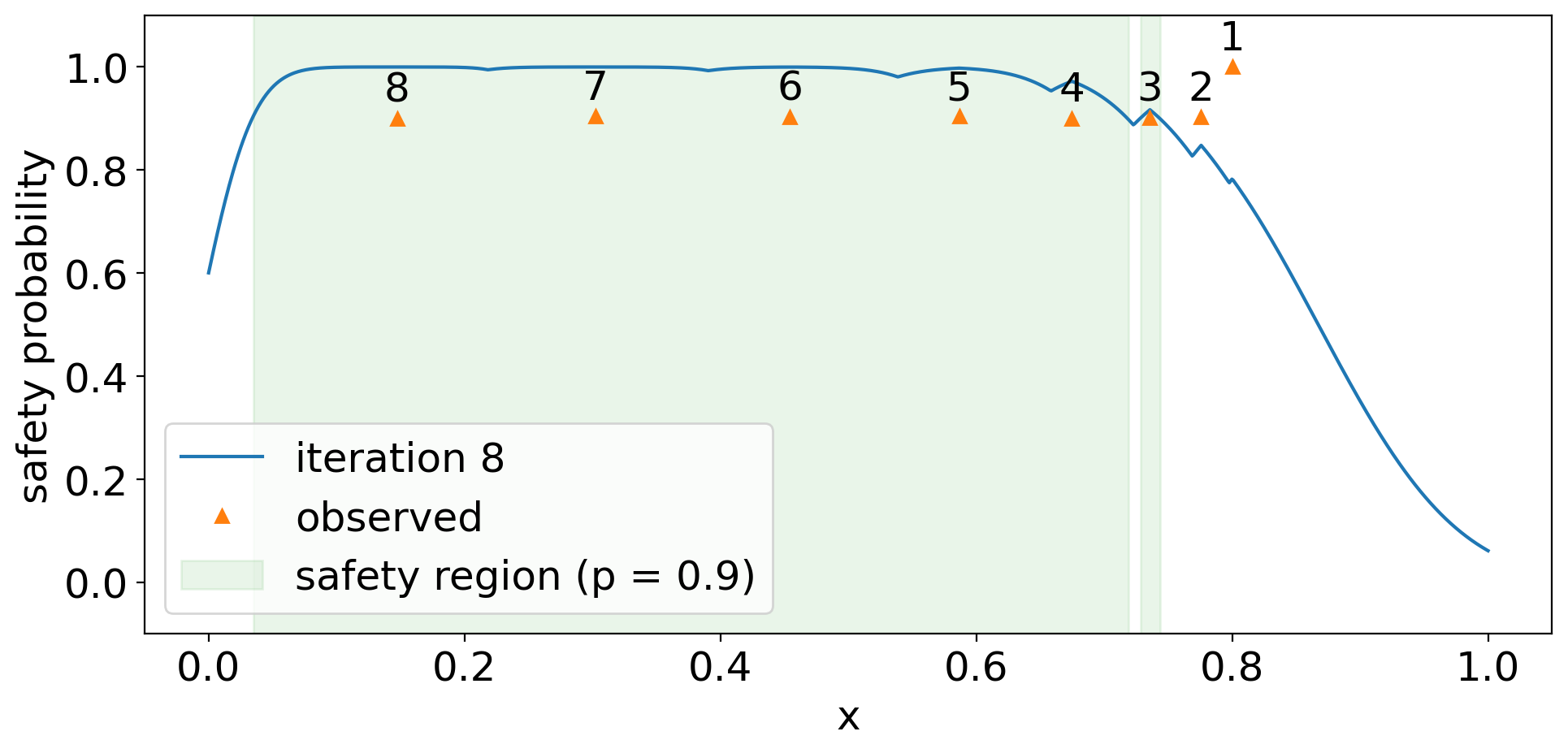}
\caption{Snapshot of the safety probability along the parameter range at the end of 1-D safety exploration. Safety probability curve is calculated based on the Gaussian random walk drift model with the previous observations (orange triangles). 
The order of sampling is denoted by the number on top of each observation.
Blue curve is the calculated safety probability and the green shaded area
shows the safety region,
with the safety probability threshold $p_s=0.9$.
}\label{fig:safety_prob_drift_a}
\end{center}
\end{figure}

\begin{figure}[htbp]
\begin{center}
\includegraphics[width=\linewidth]{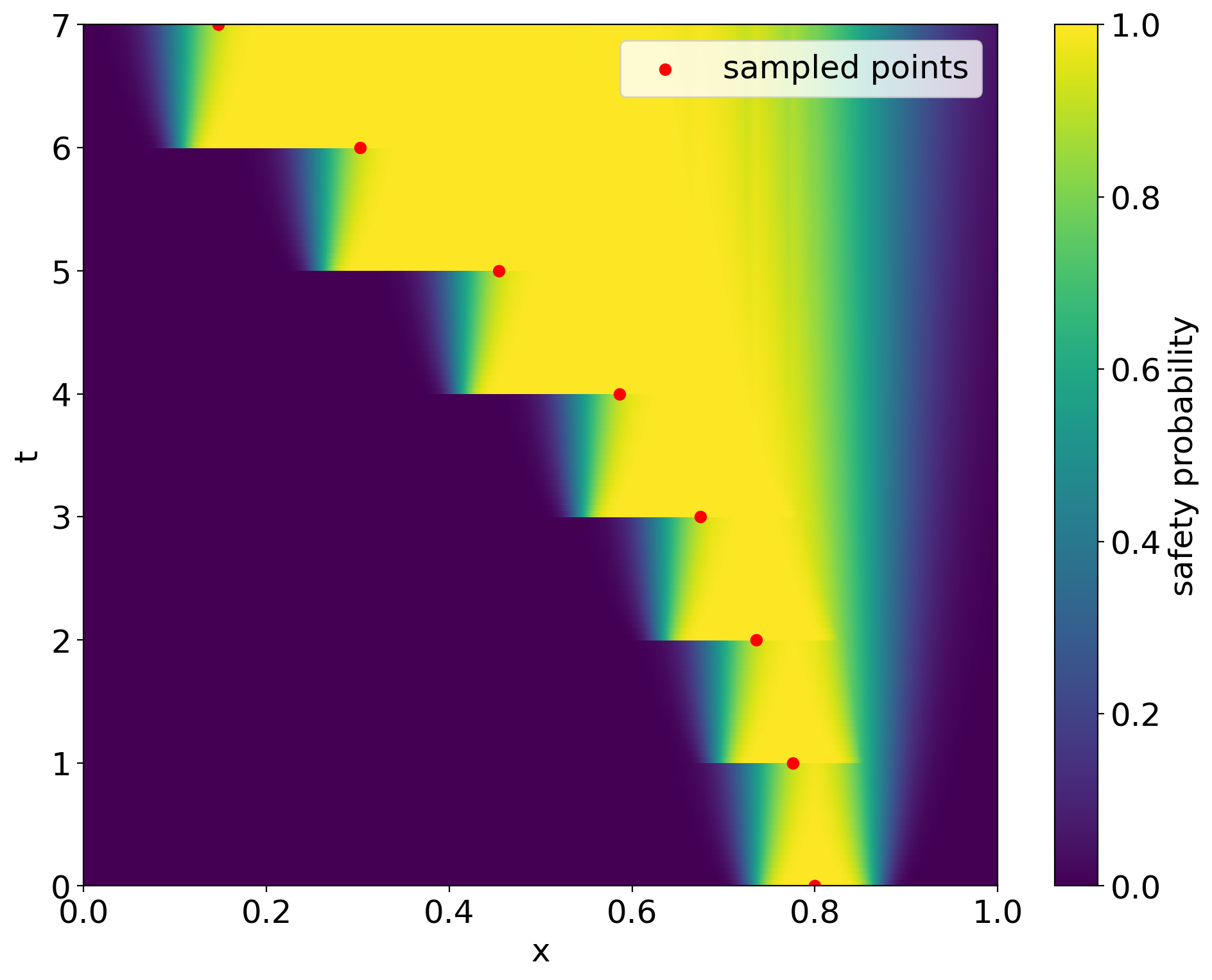}
\caption{Evolution of the safety probability during a safety exploration on the drifting 1-D test problem. Red dot denotes the time and position of the corresponding sampled point. The safety probability is calculated with the Gaussian random walk drift model.
}\label{fig:drift_model}
\end{center}
\end{figure}

Fig.~\ref{fig:safety_prob_drift_a} is the snapshot of the safety probability at the end of the safety exploration.
Clearly, the calculated safety probability curve is altered significantly compared to the case without drift (see Fig.~\ref{fig:safety_prob}).
The observation points (orange triangles) to the right side are older samples, which have less impact on the safety curve, as expected. 
The evolution of the safety probability over the test problem domain during the safety exploration
is shown in Fig.~\ref{fig:drift_model}.

\subsection{The RCDS-S algorithm}

Combining the scheme of iterative 1-D optimization over conjugate directions~\cite{Powell},  the safety 1-D exploration discussed in previous sections and the use of parabolic fitting to determine the minimum~\cite{HUANG201377}, we arrive at the RCDS-S algorithm. The algorithm logic flow is provided in Algorithm~\ref{alg:rcdss}. In the algorithm implemenation, we choose to normalize all parameters to within the range of [0, 1]. 
In the following we discuss two aspects regarding the application of the method in more details. 

First, after each iteration, when the algorithm has gone through all directions, one could replace one of the existing directions with the new direction that goes from the previous minimum to the new minimum. In the traditional Powell method, this decision is done by assessing an additional point on the new direction and using the result to estimate the potential gain from the direction~\cite{PressNumerRecipt3rd}. 
However, the additional evaluation could be unsafe. As a simple solution, the RCDS-S algorithm provides an option to always replace or not replace the direction. The option of not to replace the direction is expected to be more suitable in cases where a conjugate direction set is available through model calculation or analysis of past measurement data. 

\IncMargin{0.4em}
\begin{algorithm}
\SetAlgoLined
 $i \gets 0$, $n \gets 0$. Initialize the conjugate direction set $\mathcal{M}\in\mathbb{R}^{m\times m}$, initial solution $\mathbf{x}_0\in\mathbb{R}^m$. Define $\mathbf{v}_k \coloneqq \textrm{col}_k(\mathcal{M})$\;
 \tcc{$I_{\max}$ and $N_{\max}$ are maximum number of iterations and evaluations, respectively}
 \While{$i < I_{\max}$ \textbf{and} $n < N_{\max}$}{
   $d \gets 0$, $\eta \gets 0$\;
   \For{$k = 1$ \KwTo $m$}
   {
     Perform safety exploration at $\mathbf{x}_{k-1}$ along $\mathbf{v}_k$ to find extremum $\mathbf{x}_k$\;
     \If{$y_{k-1} - y_k > d$} {
       $d \gets y_{k-1} - y_k$\;
       $\eta \gets k$\;
     }
     $n \gets n + \textrm{number of evaluations in safety exploration}$\;
   }
   \uIf{replace direction} {
     $\mathbf{v}_{m+1} \gets \textrm{unit vector along } \mathbf{x}_m - \mathbf{x}_0$\;
     Perform safety exploration at $\mathbf{x}_m$ along $\mathbf{v}_{m+1}$ to find extremum $\mathbf{x}_{m+1}$\;
     $n \gets n + \textrm{number of evaluations in safety exploration}$\;
     $\mathbf{x}_0 \gets \mathbf{x}_{m+1}$\;
     \tcc{discarding the direction of largest decrease}
     \For{$k = \eta$ \KwTo $m$}
     {
       $\mathbf{v}_k \gets \mathbf{v}_{k+1}$\;
     }
   }
   \Else{
     $\mathbf{x}_0 \gets \mathbf{x}_m$\;
   }
   $i \gets i + 1$\;
 }
 \caption{RCDS-S}
 \label{alg:rcdss}
\end{algorithm}
\DecMargin{0.4em}

The second important aspect is about the choice of the hyper-parameters for safety exploration. 
Two most critical hyper-parameters in the modeling of the safety probability are the Lipschitz constant $L_v$ in normalized decision space and the drift rate $\sigma_d$ in time. Choosing good values for these two parameters is crucial to the safety performance of the proposed algorithm. Compared to the ideal value, a smaller $L$ would result in less safe exploration by increasing the chance of sampling an unsafe point. On the other hand, a larger $L$ encourages more conservative exploration that wastes more points for each direction and would lead to slower convergence.
Slow convergence is problematic, especially for the drifting case, as it can cause the algorithm to fail to follow the drifting optimum.

The hyper-parameters can be obtained before applying the algorithm to a specific problem by analyzing historic data or performing additional measurements. 
For example, the  maximum drifting rate $\sigma_d$ can be estimated by observing the variation of the objective function in a long period of time when no knobs are varied.
The Lipschitz $L$ parameter can be determined through 
conjugate direction scans, which are to perform full-range linear scans over the directions in the initial conjugate direction set of the normalized decision space. The maximum absolute gradient of all the scanned directions can be used as the  $L_v$ parameter. 
In the algorithm we use one $L$ parameter for all directions. In the case when no direction is replaced during execution of the algorithm, a different $L$ value could be provided for each direction. The more accurate hyper parameter could lead to better safety and efficiency. 

\begin{figure}[htbp]
\begin{center}
\includegraphics[width=\linewidth]{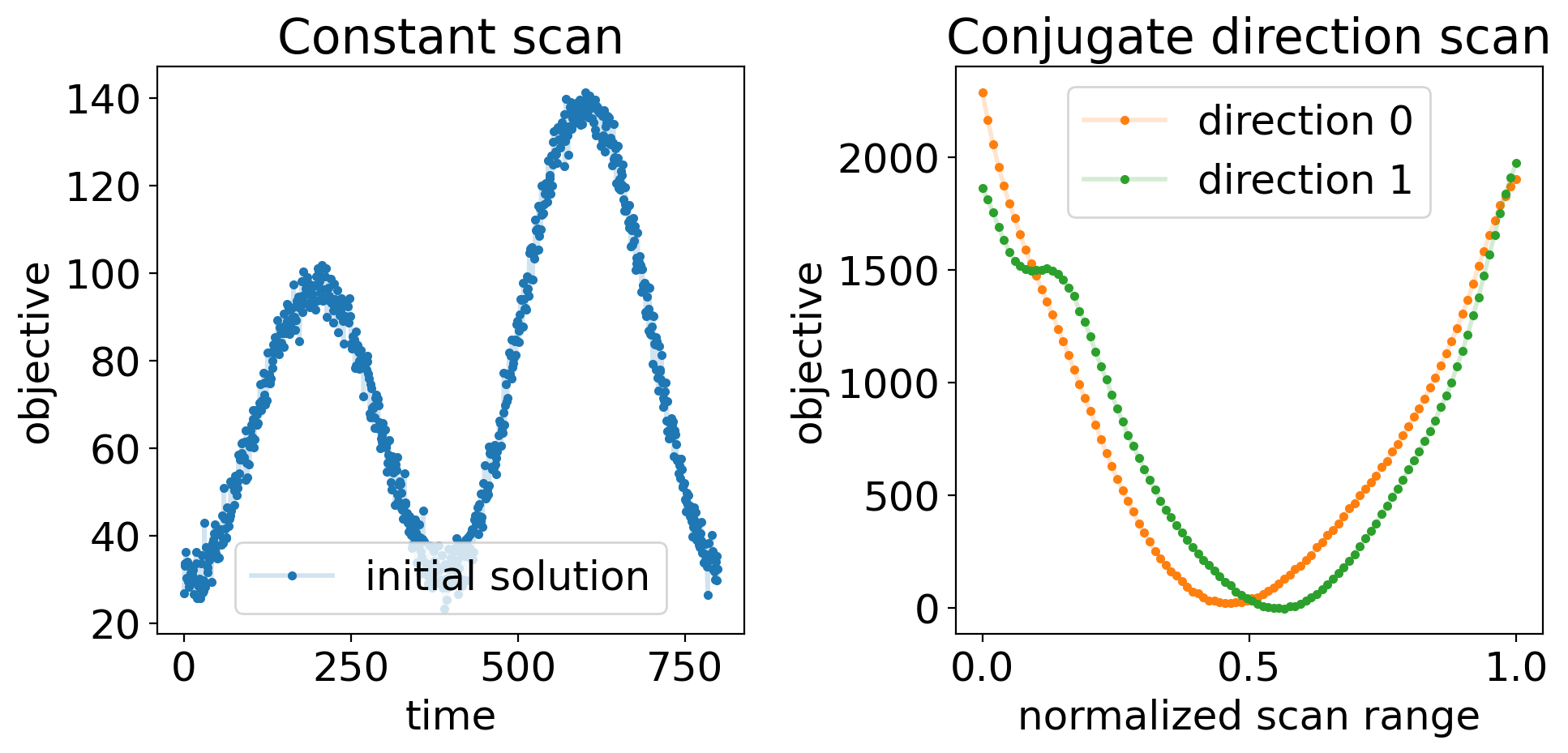}
\caption{Determination of the hyper-parameters for the kicker bump matching test problem. 
Left: variation of objective function over time; right: objective function vs. parameter scan in two directions. 
The initial solution is set to be the middle point in the 2-D normalized decision space $(0.5, 0.5)$. The estimations of $L_v$ and $\sigma_d$ are 4000 and 0.8 respectively, derived from the scan results.
}
\label{fig:two_scans}
\end{center}
\end{figure}

Because the RCDS-S algorithm relies on probability models to choose trial solutions and  these models strongly depend on the accuracy of hyper-parameters, we do not expect all trial solutions are guaranteed to be safe. There could be times when the algorithm fails to follow the drift. In other times, even when the algorithm is following the drift, there could be occasional trial solutions that exceed the safety threshold.
Therefore, while the algorithm tends to allow safe exploration of the parameter space, it is not recommended for high risk applications. 

\section{Simulation Tests\label{secSimul}}

Simulation studies were conducted to test the performance of the RCDS-S algorithm in online 
optimizations for drifting problems.
Two simulated accelerator test problems are used. 
The first  problem is kicker-bump matching for a storage ring, with a 2-D
decision space. The second one is the steering of the injected beam in the booster to storage ring (BTS) transport line to maximize injection efficiency~\cite{HUANG201377}, which is a 4-D problem.

\subsection{Kicker-bump matching safety optimization}

The Stanford Positron and Electron Asymmetric Ring-III (SPEAR3) storage ring has three injection kickers that, ideally, form a closed orbit bump when fired. In reality, the kicker bump is not strictly closed, leading to residual oscillations after the kickers are fired. 
The purpose of the kicker bump matching program is to find the kicker setting that minimize the residual oscillations. 

In the test, the strength of one kicker, K1, 
is modulated in a sinusoidal form to simulate the systematic drift, as shown in \begin{equation}
v(t) = v_0 + \sigma_d\left[\sin\left(\frac{2\pi}{p}(t + t_0)\right)-\sin\left(\frac{2\pi}{p}t_0\right)\right],
\label{eq:modulate}
\end{equation}
where $p$ is the drifting period, $\sigma_d$ the drifting amplitude, $t_0$ the time origin, $v_0$ the initial value of the modulated variable.

The strengths of the other two kickers, K2 and K3, are used as tuning knobs of the optimization problem.
The rms turn-by-turn horizontal orbit from 256 turns of residual oscillations is used as the objective function. 
A Gaussian noise with strength $\sigma_n$ is added to the objective. The problem is  normalized so that the knob values vary between 0 and 1.

For the algorithm tests, we set 
$p=800$, $\sigma_n=3$~$\mu$m, $\sigma_d=0.1$ $\mu$m per evaluation period. 
Based on the two pre-scans discussed in the last section, the Lipschitz constant $L$ and strength of Gaussian random walk $\sigma_g$ are chosen to be 2000 and 0.2, respectively. 
The safety threshold $h$ can be varied to change the safety search difficulty. In the tests, the safety threshold is set to 40 $\mu$m, which is only slightly higher than most of the observed values of the noisy objective at the initial solution.

\begin{figure}[htbp]
\begin{center}
\includegraphics[width=\linewidth]{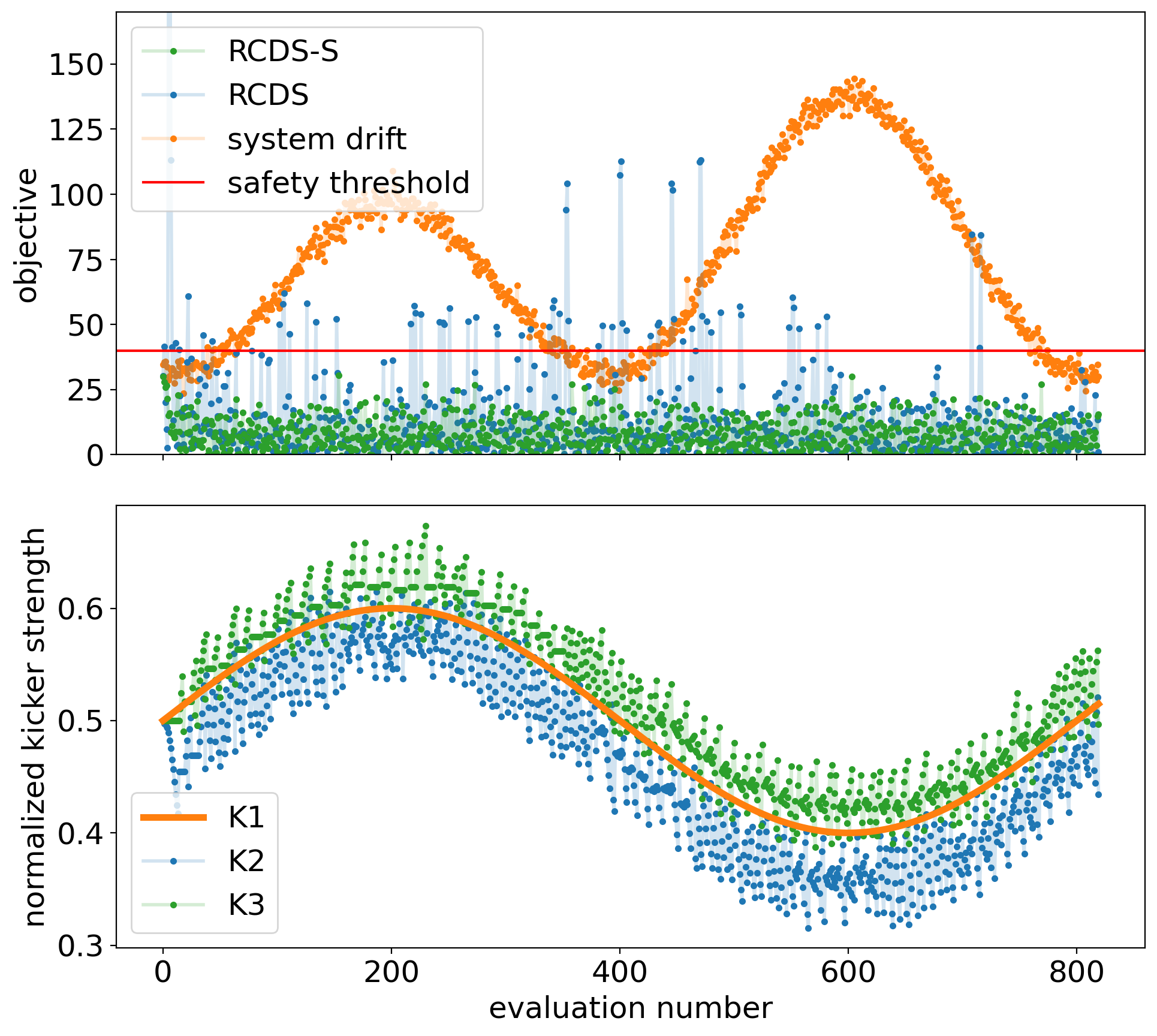}
\caption{RCDS-S for kicker-bump matching in simulation. Top plot: orange dots show the objective drift caused by the modulated kicker strength at the initial solution. Blue dots denote the objective function during optimization by the  RCDS algorithm, while green dots show the performance of RCDS-S. Red horizontal line shows the safety threshold.
Bottom plot: orange curve shows the modulation on K1, the other two dotted curves visualize the evolution history of the variables tuned by RCDS-S.
}\label{fig:kbm_sim}
\end{center}
\end{figure}

The tests have been run multiple times, and the performance is stable. A typical test  is shown in Fig.~\ref{fig:kbm_sim}. The top plot compares the objective function over one modulation period for three cases: no optimization,
tuning with RCDS,
and tuning with RCDS-S.
The test result shows that RCDS-S is able to follow the drift and seek the optimum while keeping the objectives of the trial solutions well below the safety threshold. RCDS is also very efficient for the test problem. However, since it is not aware of the safety threshold, the proposed solutions are not guaranteed to be safe. 

\subsection{BTS injection efficiency safety optimization}
The RCDS-S algorithm has also been tested with a 4-D problem in simulation. The test problem is to maximize the injection efficiency with steering knobs in the BTS transport line. 
Two pairs of upstream correctors, one pair in each transverse plane,  are modulated according to Eq.~\eqref{eq:modulate} to simulate systematic drift.
Two pairs of correctors, also one pair for each plane, toward the end of transport line are used as tuning knobs to compensate the trajectory drift and its impact to injection efficiency. 
The objective function is the negated injection efficiency. The noise level $\sigma_n$ is set to {$0.05$} and drift period $p=800$.
The drift amplitude is large enough such that the injection efficiency drops to zero for half of the modulation period. 
The safety threshold is set to $-0.5$ (i.e., corresponding to a 50\%  injection efficiency).

\begin{figure}[htbp]
\begin{center}
\includegraphics[width=\linewidth]{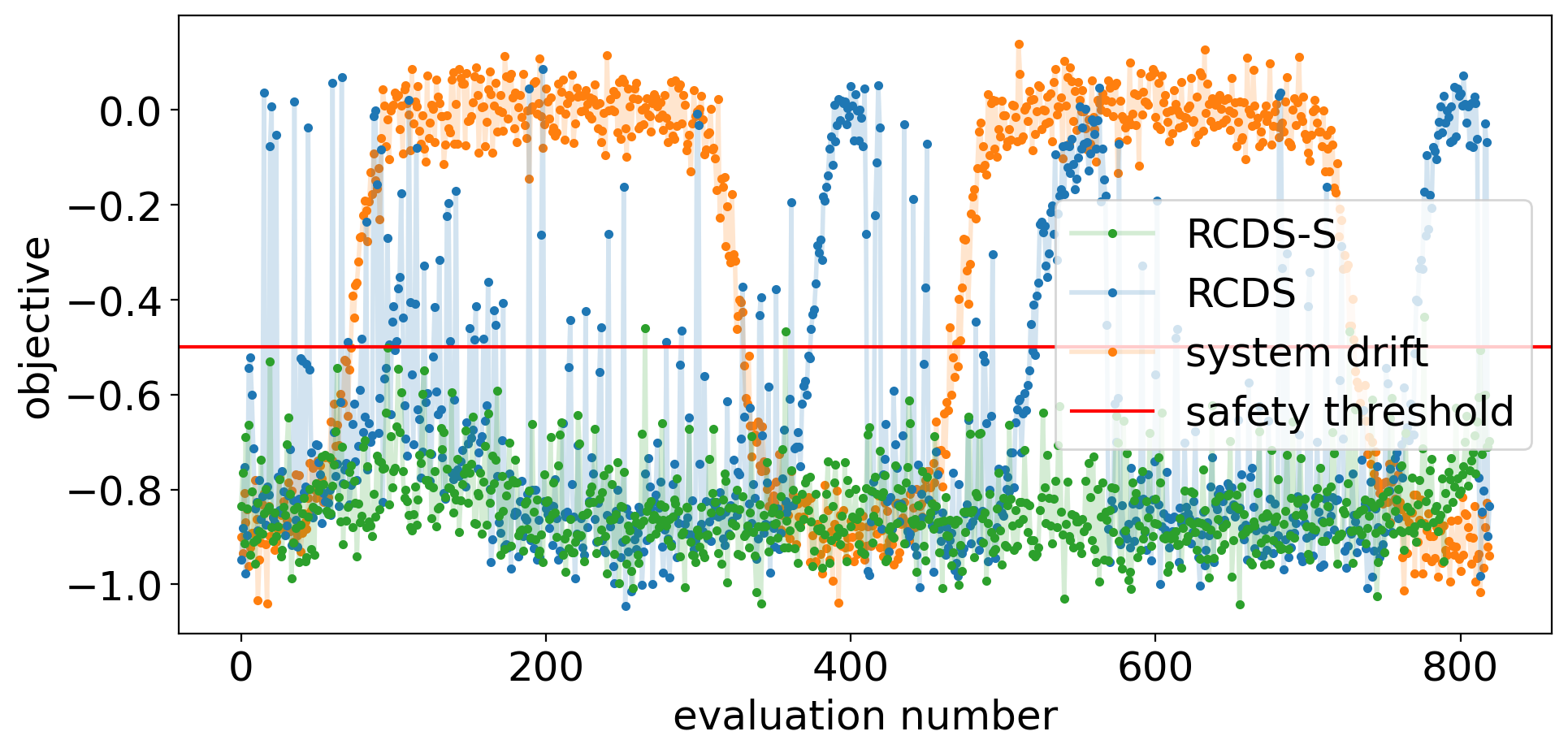}
\caption{
RCDS-S on BTS steering optimization. The objective function is the negated injection efficiency. The orange, blue, and red curves show the cases without tuning, tuning with RCDS, or tuning with RCDS-S, respectively. 
}\label{fig:bts_sim}
\end{center}
\end{figure}

Fig.~\ref{fig:bts_sim} shows the results of a typical test run.
The RCDS-S algorithm keeps the injection efficiency near the maximum and its trial solutions are mostly within the safe region. In comparison, the RCDS algorithm not only exceeds the safety threshold, but also occasionally fails to follow the drifting.

\section{Experimental application of the RCDS-S algorithm\label{secExpe}}

The RCDS-S method has been applied experimentally to the  kicker bump matching problem on the SPEAR3 storage ring
to demonstrate its capability to compensate drifts in online optimization.
The setup of the problem is the same
as was discussed in the previous section in simulation.

The goal is to minimize the residual oscillation as seen by a turn-by-turn beam position monitor. The voltage amplitude of one kicker (K1) is modulated with a period of 800 data points, with an interval of 2 seconds between data points.  
The noise level was measured right before the algorithm test and was found to be $5$~$\mu$m. The initial solution is the same as the operation setting which  would give the best objective if the system is not drifting.

\begin{figure}[htbp]
\begin{center}
\includegraphics[width=\linewidth]{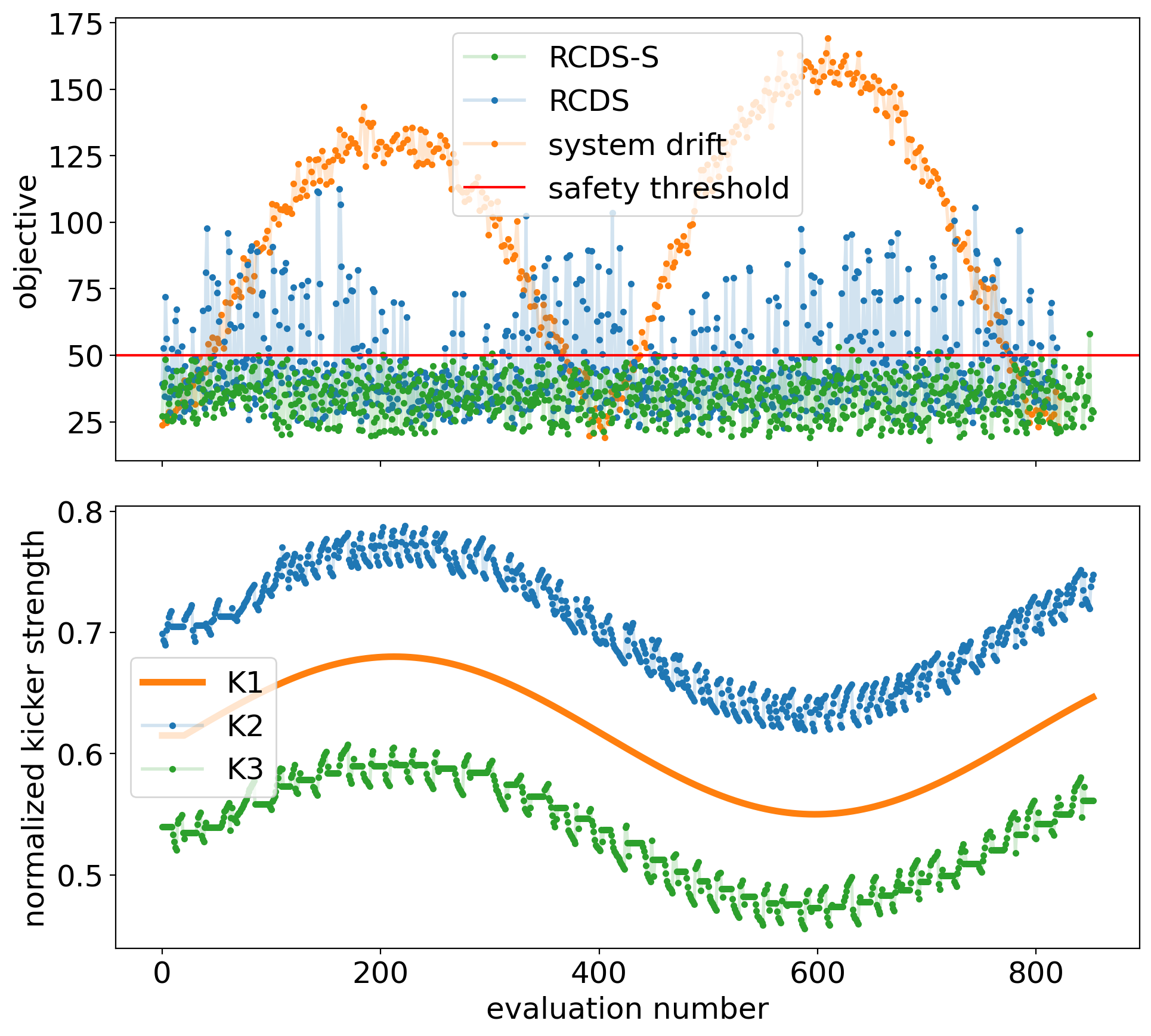}
\caption{RCDS-S on the kicker-bump matching experiment on the SPEAR3 storage ring, with safety threshold 50~$\mu$m. Top plot: orange dots show the objective drift caused by the modulated kicker strength at the initial solution. Blue dots denote the optimization performance of the  RCDS algorithm, while green dots show the performance of RCDS-S. Red horizontal line shows the safety threshold.
Bottom plot: orange curve shows the modulation on K1, the other two dotted curves visualize the evolution history of the variables as they were tuned by RCDS-S.}\label{fig:kbm_exp}
\end{center}
\end{figure}

\begin{figure}[htbp]
\begin{center}
\includegraphics[width=\linewidth]{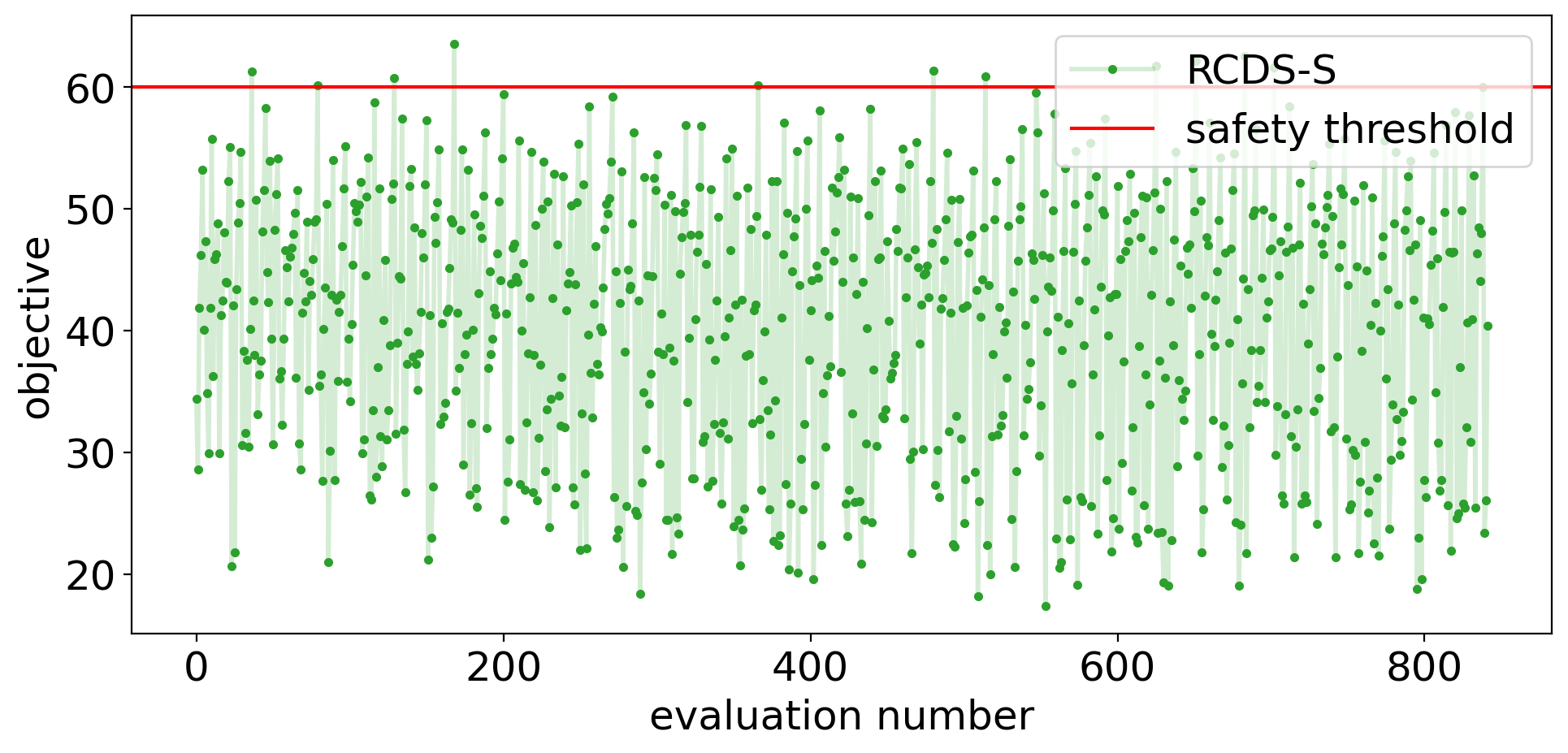}
\caption{RCDS-S with safety threshold 60 ~$\mu$m on the SPEAR3 kicker-bump matching experiment.}\label{fig:exp_cmp}
\end{center}
\end{figure}

Two rounds of tests were performed with the experimental setup. For the first round, the safety threshold is set to 60~$\mu$m.
As it succeeded, we set  the safety threshold to  50~$\mu$m for the second round. 
The results for the second round are shown in Fig.~\ref{fig:kbm_exp}, where the objective function for RCDS-S for one modulation period is compared to the results of RCDS and the case without tuning. 
The objective function for the first round is shown in Fig.~\ref{fig:exp_cmp}.

The experimental performance is similar to the simulation cases for both RCDS-S and RCDS.
The objective function has a larger  variance for the evaluated trail solutions in experiments. This could be due to the higher measurement noise level.
RCDS-S would explore until it brackets the peak of the objective curve, and hence, the higher noise level is, the wider the exploration range would be, for the 1-D exploration to gain the same level of confidence that a peak has been bracketed.

Another behavior that has been observed in experiment is that RCDS-S had a few small violation solutions for both rounds.
This could be due to a mismatch between the true $L_v$ of the experiment problem and the one we applied in the algorithm (that comes from the simulation). 
Using a larger $L_v$ could reduce the violation rate at the expense of more trial solutions per direction, and  consequently would increase the probability of RCDS-S failing to follow the systematic drift.

\section{Alternative Approach to Model Drift\label{secAlt}}

In Sec.~\ref{ssec:drift_model_a}, the  systematic drift is modeled as an increasing uncertainty with time.
With this model we are not incorporating any prior knowledge about the drift behavior of the system. It may be possible to improve the predictive capability of the model if the drift trend is included. 
In the general case, the objective function can be seen as a function of both the tuning knobs and time, 
\[
y = f(\mathbf{x}, t) + \epsilon. 
\]
Similar to the variation of knobs, one simple assumption about the dependence on time is that the rate of change has an upper limit. In this case,  in addition to the L-Lipschitz continuous condition on $\mathbf{x}$, we also assume the objective function to be $\sigma_d$-Lipschitz continuous on $t$:
\[
\left\|f(\mathbf{x}, t_1) - f(\mathbf{x}, t_2)\right\| \leq \sigma_d\cdot\left\|t_1 - t_2\right\|. 
\]
Therefore, based on the observation on a point $\mathbf{x_0}$ that was measured at time $-t$ (relative to the current time), we have:
\begin{eqnarray*}
\left\|f(\mathbf{x}, 0) - f(\mathbf{x_0}, -t)\right\| &\leq& \left\|f(\mathbf{x}, 0) - f(\mathbf{x}, -t)\right\|\\
&&+ \left\|f(\mathbf{x}, -t) - f(\mathbf{x_0}, -t)\right\|\\
&\leq& L\cdot\left\|\mathbf{x}-\mathbf{x_0}\right\| + \sigma_d\cdot t
\end{eqnarray*}
Following the same path as in the derivation of  Eq.~\eqref{equ:safety_cond}, we have:
\begin{equation}
\hat{\epsilon} \leq \frac{h - E_{\max} - \sigma_d\cdot t}{\sqrt{2\sigma_n^2}},
\label{equ:drift_b}
\end{equation}
where $E_{\max}$ is as defined previously, which is the maximum expected value at point $\mathbf{x}$ without drift.

\begin{figure}[htbp]
\begin{center}
\includegraphics[width=\linewidth]{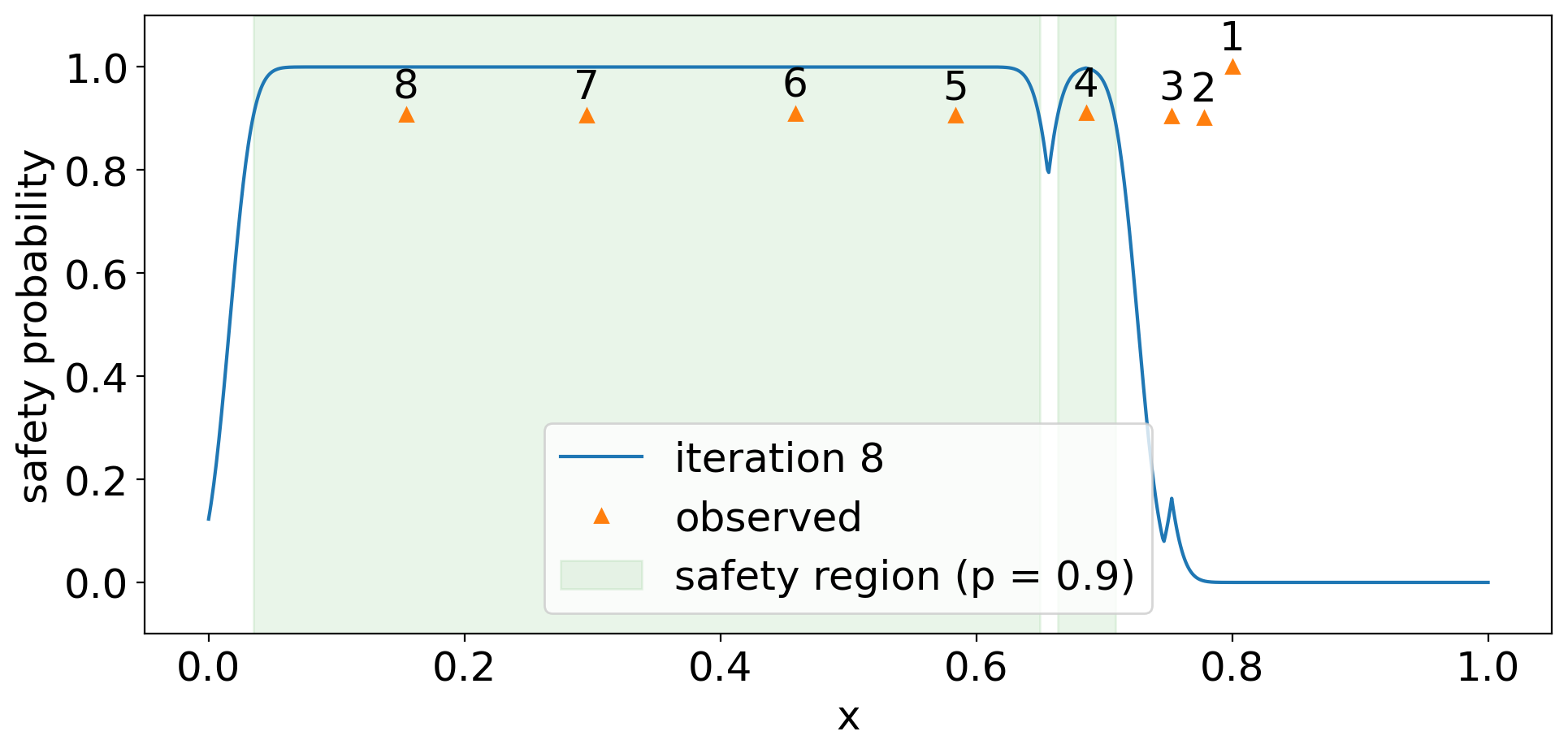}
\caption{Snapshot of the safety probability along the viable range at the end of 1D safety exploration. Safety probability curve is calculated based on the alt drift model with the given observations. Orange triangles denote the observations,
the order of sampling is denoted by the number on top of each observation.
Blue curve is the calculated safety probability and the green shaded area
shows the safety region with the safety probability threshold $p_s=0.9$.}\label{fig:safety_prob_drift_b}
\end{center}
\end{figure}

\begin{figure}[htbp]
\begin{center}
\includegraphics[width=\linewidth]{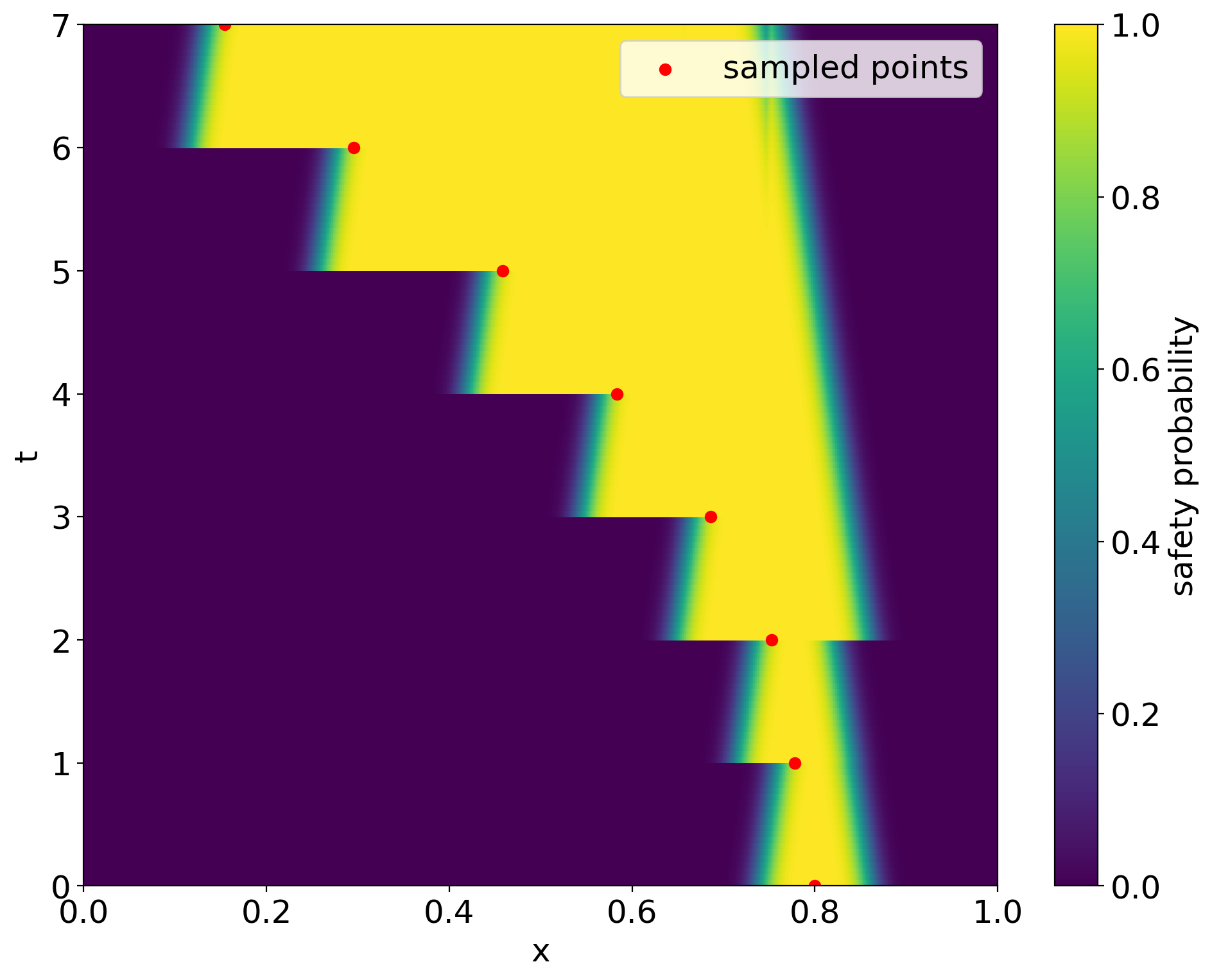}
\caption{Evolution of the safety probability during a safety exploration on the drifting 1D test problem. Red dot denotes the time and position of the corresponding sampled point. The safety probability is calculated with the alternative drift model.
}\label{fig:drift_model_alt}
\end{center}
\end{figure}

This alternative approach to perform a 1D safety exploration is applied to
the test problem discussed in Sec.\ref{sec:drift_1d}. The efficiency is found to be similar to the Gaussian random walk approach. 
However, the evolution of the safety prob curves, and consequently
the sampled points, are quite different for the two cases.
The time-sensitivity of the alternative approach is higher than the original one, i.e., the impact of the older data points to the safety prob curve decays faster, as shown in Fig.~\ref{fig:safety_prob_drift_b}. For example, the three points on the right side almost have no impact on the safety prob curve. The evolution of the safety probability during the 1D exploration is visualized in Fig.~\ref{fig:drift_model_alt}.

\section{Conclusions\label{secConclu}}
In this study, we propose an optimization algorithm (RCDS-S) that combines robust conjugate direction search and a new 1-D safety exploration algorithm to optimize noisy, drifting machine performances online, while keeping the machine performance within a designated safe envelope. 
The 1-D safety exploration algorithm makes use of Lipschitz continuity of the objective function and  properties of a Gaussian random walk process for the drift to build a probability model of the drifting function, with which to suggest safe trial solutions. 
The proposed algorithm has been successfully tested on simulated and experimental accelerator tuning problems. 

\section*{Acknowledgment}
  This work was supported by the U.S. Department of Energy, Office of
  Science, Office of Basic Energy Sciences, under Contract No.
  DE-AC02-76SF00515. 
  
\input{output.bbl}

\end{document}

%% file: output.bbl
%